\documentclass[aps,prl,10pt,twocolumn,superscriptaddress]{revtex4-1}
\usepackage{graphicx}
\usepackage{amsmath,amssymb}
\usepackage{bm}

\makeatletter
\newcommand*{\balancecolsandclearpage}{%
  \close@column@grid
  \clearpage
  \twocolumngrid
}
\makeatother

\begin{document}

\title{Spin-Fluctuation Mechanism of Anomalous Temperature Dependence of Magnetocrystalline Anisotropy in Itinerant Magnets}
\author{I.\ A.\ Zhuravlev}
\affiliation{Department of Physics and Astronomy and Nebraska Center for Materials and Nanoscience,
University of Nebraska-Lincoln, Lincoln, Nebraska 68588, USA}
\author{V.\ P.\ Antropov}
\affiliation{Ames Laboratory, U.S. Department of Energy, Ames, Iowa 50011, USA}
\author{K.\ D.\ Belashchenko}
\affiliation{Department of Physics and Astronomy and Nebraska Center for Materials and Nanoscience,
University of Nebraska-Lincoln, Lincoln, Nebraska 68588, USA}

\begin{abstract}
The origins of the anomalous temperature dependence of magnetocrystalline anisotropy in (Fe$_{1-x}$Co$_{x}$)$_{2}$B alloys are elucidated using first-principles calculations within the disordered local moment model. Excellent agreement with experimental data is obtained. The anomalies are associated with the changes in band occupations due to Stoner-like band shifts and with the selective suppression of spin-orbit ``hot spots'' by thermal spin fluctuations. Under certain conditions, the anisotropy can increase, rather than decrease, with decreasing magnetization due to these peculiar electronic mechanisms, which contrast starkly with those assumed in existing models.
\end{abstract}

\maketitle

Magnetocrystalline anisotropy (MCA) is one of the key properties of a magnetic material \cite{Stohr}. Understanding of its temperature dependence is a challenging theoretical problem with implications for the design of better materials for permanent magnets \cite{Lewis}, heat-assisted magnetic recording \cite{Kryder}, and other applications. While the MCA energy $K$ usually declines monotonically with increasing temperature as predicted by simple models \cite{CC}, in some magnets it behaves very differently and can even increase with temperature. Such anomalous $K(T)$ dependence makes some materials useful as permanent magnets and can potentially facilitate specialized applications.

Well-known anomalies in the temperature dependence of MCA include spin reorientation transitions (SRT) in cobalt \cite{Carr} and MnBi \cite{MnBi}, which have been attributed to thermal expansion; an SRT in gadolinium, which may be due to higher-order terms in MCA \cite{Brooks}; SRT in R$_2$Fe$_{14}$B hard magnets \cite{Herbst} due to the ordering of the rare-earth spins at low $T$; and SRT in thin films \cite{Pappas,Vaz} associated with the competition between the bulk and surface contributions to MCA. Competition between single-site and two-site MCA can also lead to an SRT \cite{Buruzs}.

MCA in metallic magnets is rarely dominated by the single-ion mechanism leading to the $K\propto M^3$ dependence on the magnetization \cite{CC}. For example, two-ion terms in $3d$-$5d$ alloys like FePt modify this dependence to $K\propto M^{2.1}$ \cite{Okamoto,Mryasov}. Clear understanding of the anomalous temperature dependence of MCA has been so far limited to the cases when competing contributions to MCA can be sorted out in real space, such as, for example, bulk and surface terms in thin films. In contrast, understanding of MCA in itinerant magnets usually requires a reciprocal space analysis \cite{Kondorskii}.

One such system is the disordered substitutional (Fe$_{1-x}$Co$_{x}$)$_{2}$B alloy, which exhibits three concentration-driven SRTs at $T=0$, a high-temperature SRT at the Fe-rich end, and a strongly non-monotonic temperature dependence at the Co-rich end with a low-temperature SRT \cite{Iga,APL}. The SRT's at $T=0$ were traced down to the variation of the band filling with concentration combined with spin-orbital selection rules \cite{APL}. Here we elucidate the unconventional mechanisms leading to the spectacular anomalies in the temperature dependence of MCA in this system and show that they stem from the changes in the electronic structure induced by spin fluctuations. We will see that under certain conditions MCA can increase, rather than decrease, with decreasing magnetization due to these mechanisms.

Our calculations employ the Green's function-based linear muffin-tin orbital method \cite{Turek} with spin-orbit coupling (SOC) included as a perturbation to the potential parameters \cite{APL,TDK}. Thermal spin fluctuations are included within the disordered local moment (DLM) model \cite{Oguchi,Gyorffy}, which treats them within the coherent potential approximation (CPA) on the same footing with chemical disorder. The DLM method has been previously used to calculate the $K(T)$ dependence in systems like FePt \cite{Staunton1,Staunton2} and YCo$_5$ \cite{Matsumoto}. Although $K(T)$ in these metals does not follow the Callen-Callen model \cite{CC} designed for materials with single-ion MCA, it still decreases monotonically. In contrast, we will see that the changes in the electronic structure with temperature lead to strong anomalies in (Fe$_{1-x}$Co$_{x}$)$_{2}$B. Our implementation of the DLM method is described in Ref.\ \onlinecite{FeMnPt}. (See Supplemental Material \cite{SM} for additional details.)

Apart from the inclusion of spin disorder, the computational details are similar to Ref.\ \onlinecite{APL}. In particular, the large overestimation of the magnetization in density-functional calculations for Co$_2$B (1.1 $\mu_B$ compared to experimental 0.76 $\mu_B$ per Co atom) is corrected by scaling the local part of the exchange-correlation field for Co atoms by a factor 0.8 at all concentrations. This treatment is consistent with spin-fluctuation theories showing that spin fluctuations tend to reduce the effective Stoner parameter \cite{Moriya,AS} and allows us to take into account the resulting changes in the electronic structure.

Magnetism in (Fe$_{1-x}$Co$_{x}$)$_{2}$B alloys is much more itinerant compared to systems like FePt; the spin moments of Fe and, especially, Co atoms are not rigid in density-functional calculations. To implement spin disorder within the DLM method, we make a simple assumption that the spin moments of both Fe and Co at finite $T$ can be taken from the ferromagnetic state at $T=0$. This assumption is based on the expectation that thermal spin fluctuations to a large extent restore the ``soft'' spin moments \cite{Moriya}. On the other hand, the variation of the electronic structure with $T$ should not be very sensitive to the details of the spin fluctuation model. For simplicity, a similar approach is used for the (Co$_{1-x}$Ni$_{x}$)$_{2}$B system, including the small spin moments on the Ni atoms.

The distribution functions for spin orientations are taken in the Weiss form: $p_\nu(\theta)\propto\exp(\alpha_\nu\cos\theta)$, where $\theta$ is the angle made by the spin with the magnetization axis, and $\nu$ labels the alloy component. The temperature dependence of the coefficients $\alpha_\nu$ is determined using the calculated effective exchange parameters, as explained in the Supplemental Material \cite{SM}. Fermi-Dirac smearing is neglected, because the effects of spin fluctuations are overwhelmingly stronger.

\begin{figure}[htb]
\includegraphics[width=0.9\columnwidth]{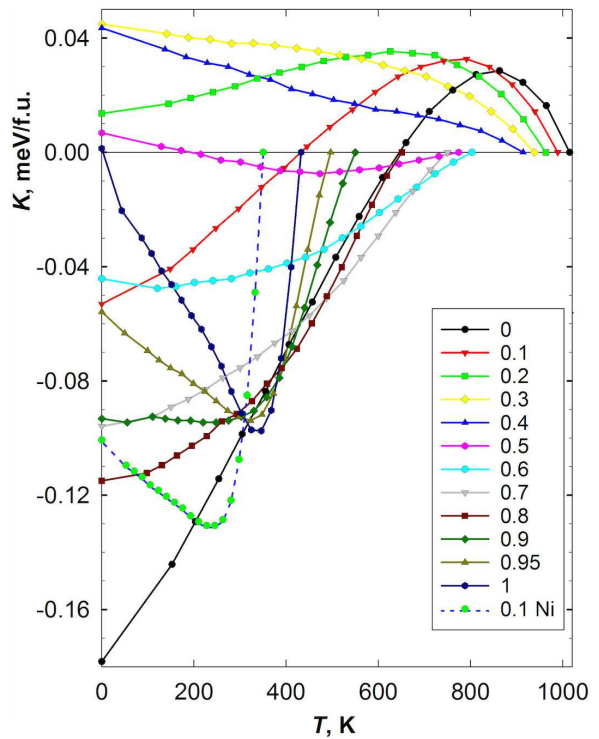}
\caption{Calculated temperature dependencies of MCA energy  $K$ in (Fe$_{1-x}$Co$_x$)$_2$B and (Co$_{0.9}$Ni$_{0.1}$)$_2$B alloys.}
\label{maetempdep}
\end{figure}

The results of $K(x,T)$ calculations shown in Fig.\ \ref{maetempdep}, which were obtained with temperature-independent lattice parameters, are in excellent agreement with experimental data \cite{Iga}. Both the thermal SRT at the Fe-rich end and the non-monotonic temperature dependence at the Co-rich end in (Fe$_{1-x}$Co$_{x}$)$_{2}$B alloys are captured (see Supplemental Material \cite{SM} for a direct comparison). For (Co$_{0.9}$Ni$_{0.1}$)$_{2}$B the MCA energy at $T=0$ is large and negative in agreement with experiment \cite{Iga}, although the initial decline of $K(T)$ similar to Co$_2$B is not observed in experiment. The finite slope in the $K(T)$ curves at zero temperature is due to the classical treatment of spin fluctuations. We have explicitly verified that the effect of thermal expansion on $K(T)$ in Fe$_2$B and Co$_2$B is almost unnoticeable.

\begin{figure}[hbt]
\includegraphics[width=0.9\columnwidth]{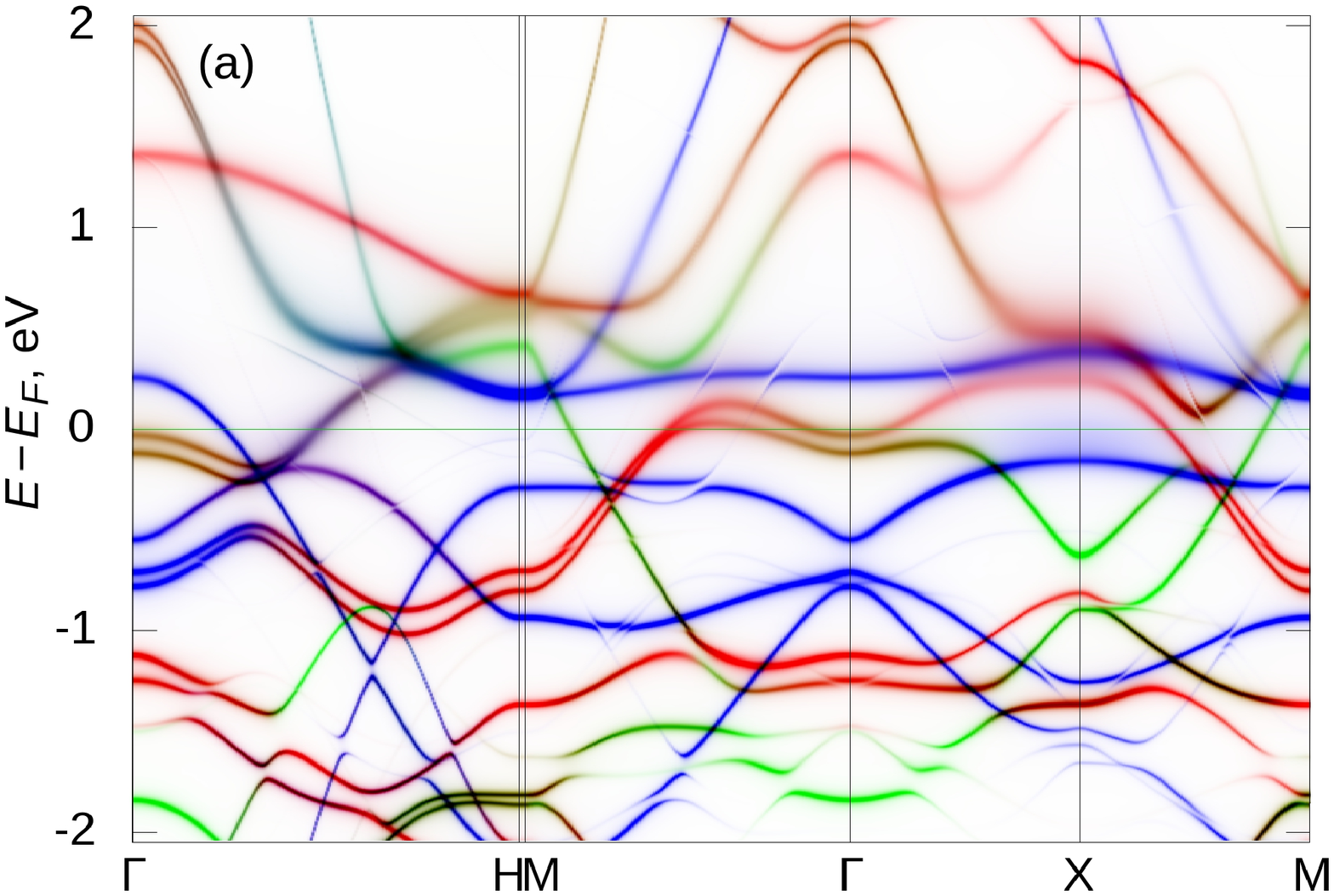}\vskip2ex
\includegraphics[width=0.9\columnwidth]{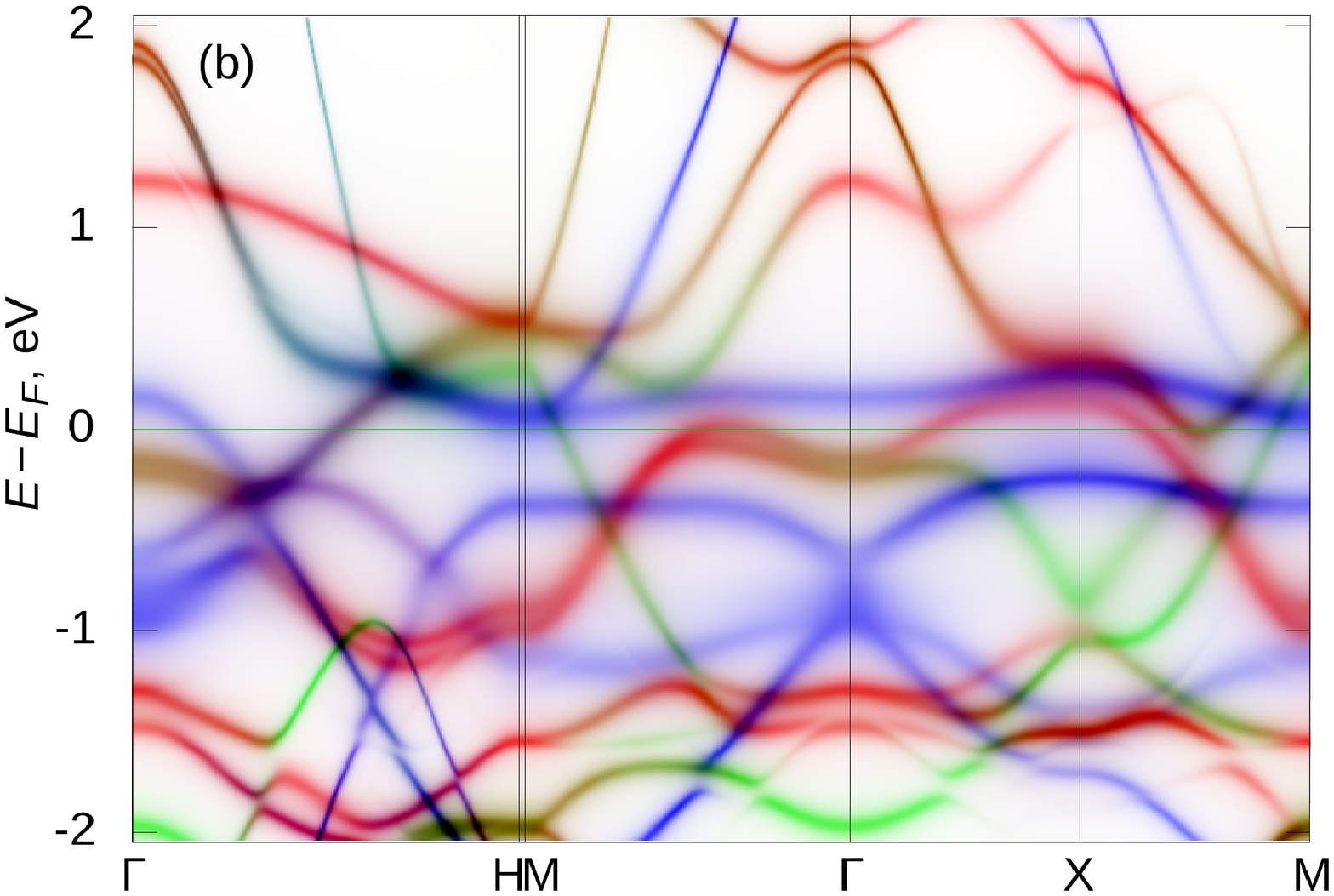}
\caption{Partial minority-spin spectral function for the transition-metal site in (Fe$_{0.05}$Co$_{0.95}$)$_2$B at (a) $T=0$, and (b) $T/T_C=0.7$. SOC is included, $\mathbf{M}\parallel z$, and energy is in eV. Color encodes the orbital character of the states. The intensities of the red, blue and green color channels are proportional to the sum of $m=\pm2$ ($xy$ and $x^2-y^2$), sum of $m=\pm1$ ($xz$ and $yz$), and $m=0$ ($z^2$) character, respectively.}
\label{fig:sf}
\end{figure}

The effects of spin disorder on the electronic structure can be understood from Fig.\ \ref{fig:sf}, which shows the partial minority-spin Bloch spectral function at $x=0.95$ for $T=0$ and $T/T_C=0.7$. Here, at the Co-rich end, all bands are easily identifiable and relatively weakly broadened at $T/T_C=0.7$. In addition, they are shifted down relative to their positions at $T=0$, which is a hallmark of an itinerant Stoner system. In contrast, at the Fe-rich end the bands are strongly broadened by spin fluctuations, so that most bands in the 1 eV window below $E_F$ are barely visible (see Supplemental Material \cite{SM}). The large difference in the degree of band broadening between the Fe-rich and Co-rich ends is due to the 2.5-fold difference in the magnitude of the spin moments. The effect of phonon scattering on band broadening in (Fe$_{1-x}$Co$_{x}$)$_{2}$B alloys is likely much smaller and is neglected here.

The usual expectation is that spin disorder should reduce MCA as a result of averaging over spin directions. Such normal behavior is seen, for example, at $x=0.3$ in Fig.\ \ref{maetempdep}. This expectation is violated at many concentrations: $K(x,T)$ is non-monotonic with respect to $T$ at $0\leq x\leq0.2$, $0.5\leq x \leq0.6$, and $0.9\leq x\leq1$; we will call this behavior anomalous. At $x\leq 0.6$ the anomalous temperature dependence of $K$ at a given $x$ follows the variation of $K$ with increasing $x$ at $T=0$. For example, $K(0.2,0)>K(0.1,0)$, and $K(0.1,T)$ anomalously increases with $T$. At $x\geq0.9$ the anomalous variation is opposite to the trend in $K(x,0)$ with increasing $x$. To understand this difference, we first need to examine the effect of disorder on MCA.

\begin{figure}[htb]
\includegraphics[width=0.9\columnwidth]{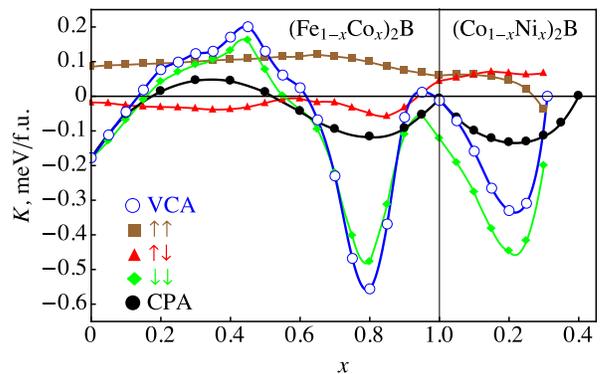}
\caption{MCA in (Fe$_{1-x}$Co$_x$)$_2$B and (Co$_{1-x}$Ni$_x$)$_2$B alloys calculated within VCA (empty circles) compared with CPA (filled circles). The spin decomposition is given for VCA.}
\label{MAEvca}
\end{figure}

Fig.\ \ref{MAEvca} compares $K(x,0)$ calculated within the virtual crystal approximation (VCA) with CPA results for (Fe$_{1-x}$Co$_x$)$_2$B \cite{APL} and (Co$_{1-x}$Ni$_x$)$_2$B systems \cite{Ni-note}. Note that in the (Co$_{1-x}$Ni$_x$)$_2$B system the spin moments vanish near 40\% Ni, in agreement with experiment \cite{Buschow}. In addition to the MCA energy $K$, Fig.\ \ref{MAEvca} also shows its approximate spin decomposition $K_{\sigma\sigma'}$ obtained from the SOC energy \cite{APL,SM}. Because the $3d$ shell in this system is more than half filled, the variation of MCA with $x$ is largely controlled by the $K_{\downarrow\downarrow}$ term, i.e., by the $L_zS_z$ mixing of the minority-spin states. Substitutional disorder strongly suppresses MCA, an effect that was also found in tetragonal Fe-Co alloys \cite{Turek2012}. The suppression is due to band broadening, which reduces the efficiency of spin-orbital selection rules. Importantly, bands broaden at different rates; the contributions to MCA from the bands that lie close to $E_F$ and broaden strongly are most effectively suppressed. The dispersive majority-spin bands are weakly broadened, and hence the $K_{\uparrow\uparrow}$ term is almost unaffected by disorder; in contrast, $K_{\downarrow\downarrow}$ is strongly reduced. We note that although band broadening (and thereby MCA) can depend on chemical short-range order, the latter is expected to be negligible in the present alloy with chemically similar constituents.

\begin{figure}[htb]
\centering
\includegraphics[width=0.9\columnwidth]{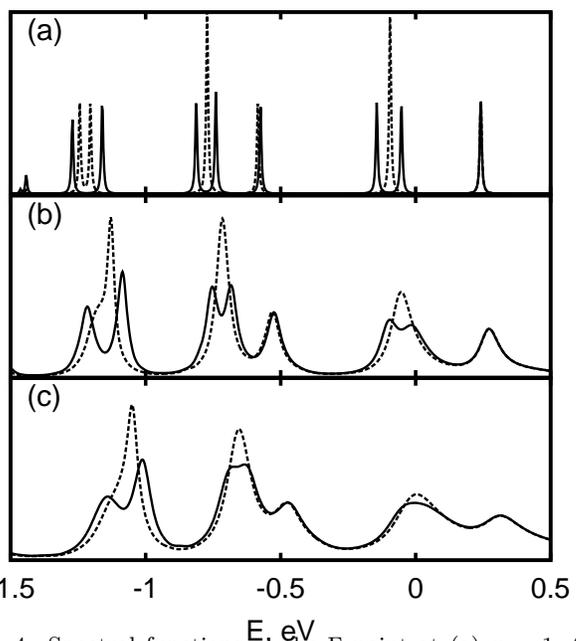}
\caption{Spectral functions at the $\Gamma$ point at (a) $x=1$, (b) $x=0.9$, (c) $x=0.8$. Solid lines: $\mathbf{M}\parallel z$. Dashed lines: $\mathbf{M}\parallel x$. A small imaginary part is added to energy to resolve the bands in panel (a).}
\label{splitting}
\end{figure}

The strongest suppression of MCA can be expected for the ``hot spots'' appearing when nearly degenerate bands at $E_F$ are split by SOC \cite{Kondorskii}. A clear example of such bands is seen near the $\Gamma$ point in Fig.\ \ref{fig:sf}a. The effect of disorder is further illustrated in Fig.\ \ref{splitting} showing the spectral function at the $\Gamma$ point for two orientations of the magnetization at $x=1$, 0.9, and 0.8, all at $T=0$. At $x=1$ there is no disorder, and the sharp bands are fully split by SOC for $\mathbf{M}\parallel z$. With the addition of Fe, the broadening quickly exceeds the original SOC-induced splitting, and the effect of SOC is strongly suppressed.

Disorder has a similar effect on the mixing of electronic bands of opposite spin by $L_+S_-$ and $L_-S_+$. Indeed, while in Fig.\ \ref{fig:sf}a for $T=0$ the anticrossings with the majority-spin bands are clearly visible, in Fig.\ \ref{fig:sf}b, for $T/T_C=0.7$, they are almost completely suppressed.

We now return to the analysis of the anomalous temperature dependence of $K$. We expect that these anomalies come from the effects of thermal spin fluctuations on the electronic structure beyond a simple averaging over spin directions. As we saw in Fig.\ \ref{fig:sf}, there are two such effects in (Fe$_{1-x}$Co$_x$)$_2$B: reduction of the exchange splitting $\Delta$, and band broadening. The reduction of $\Delta$ shifts the minority-spin bands downward relative to $E_F$, just as the band filling with increasing $x$ does. Band broadening has a stronger effect on the minority-spin states, where $E_F$ lies within the relatively heavy $3d$ bands, and it is particularly important for nearly degenerate bands straddling the Fermi level, as we saw in Fig.\ \ref{splitting}.

To understand how these effects lead to to the anomalies in $K(T)$, it is convenient to examine two quantities, $K_\uparrow$ and $K_\downarrow$, defined as $K_\sigma=\int^{E_0}(E-E_0)\Delta N_\sigma(E)dE$, where $E_0$ is the Fermi energy in the absence of SOC, and $\Delta N_\sigma$ is the difference, between $\mathbf{M}\parallel x$ and $\mathbf{M}\parallel z$, in the partial density of states for spin $\sigma$ in the global reference frame. Their sum $K_\uparrow+K_\downarrow$ closely approximates $K$, and their analysis can help identify the contributions of different bands to $K$, particularly in combination with reciprocal-space resolution \cite{APL,SM}.

\begin{figure}[htb]
\includegraphics[width=0.9\columnwidth]{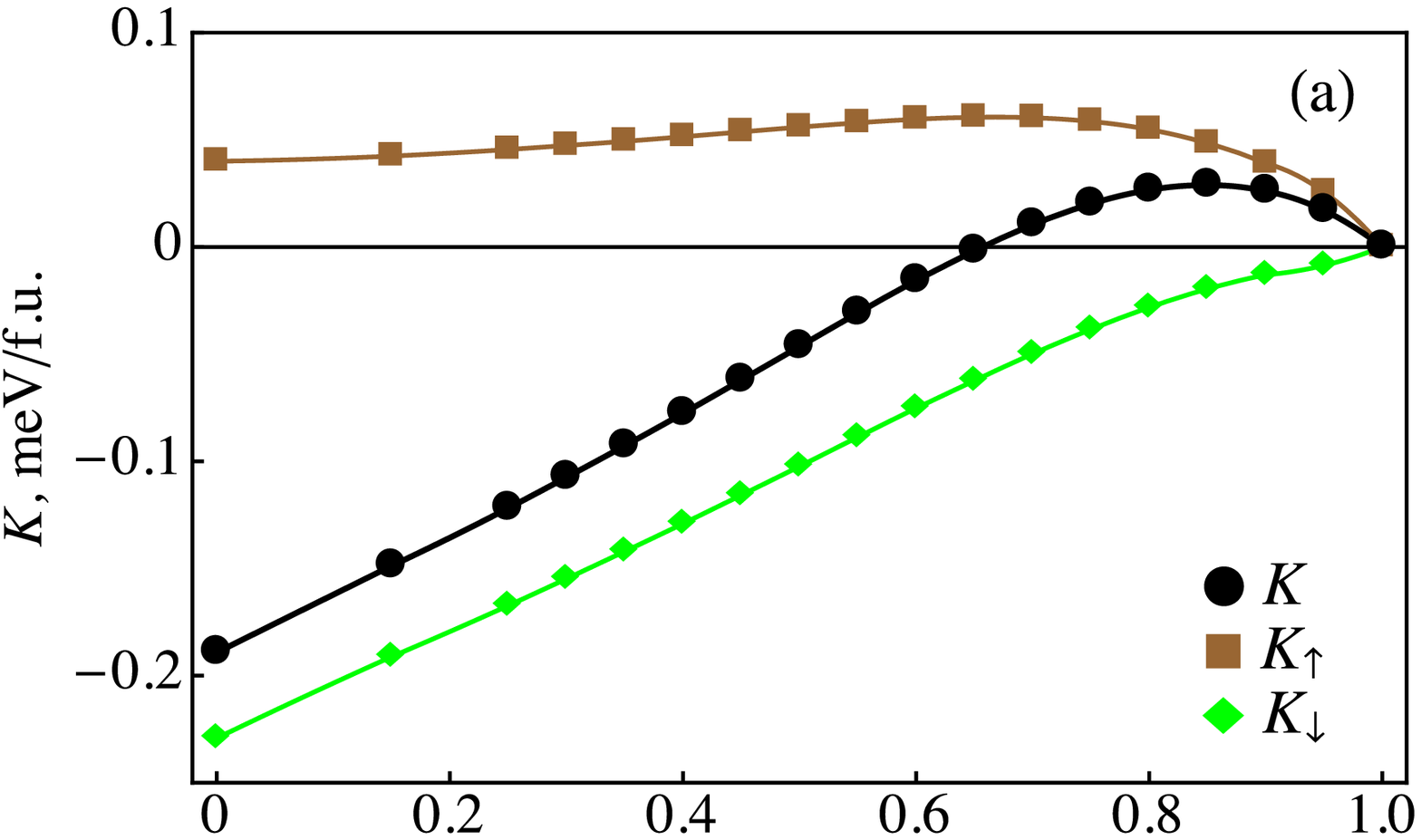}\vskip2ex
\includegraphics[width=0.9\columnwidth]{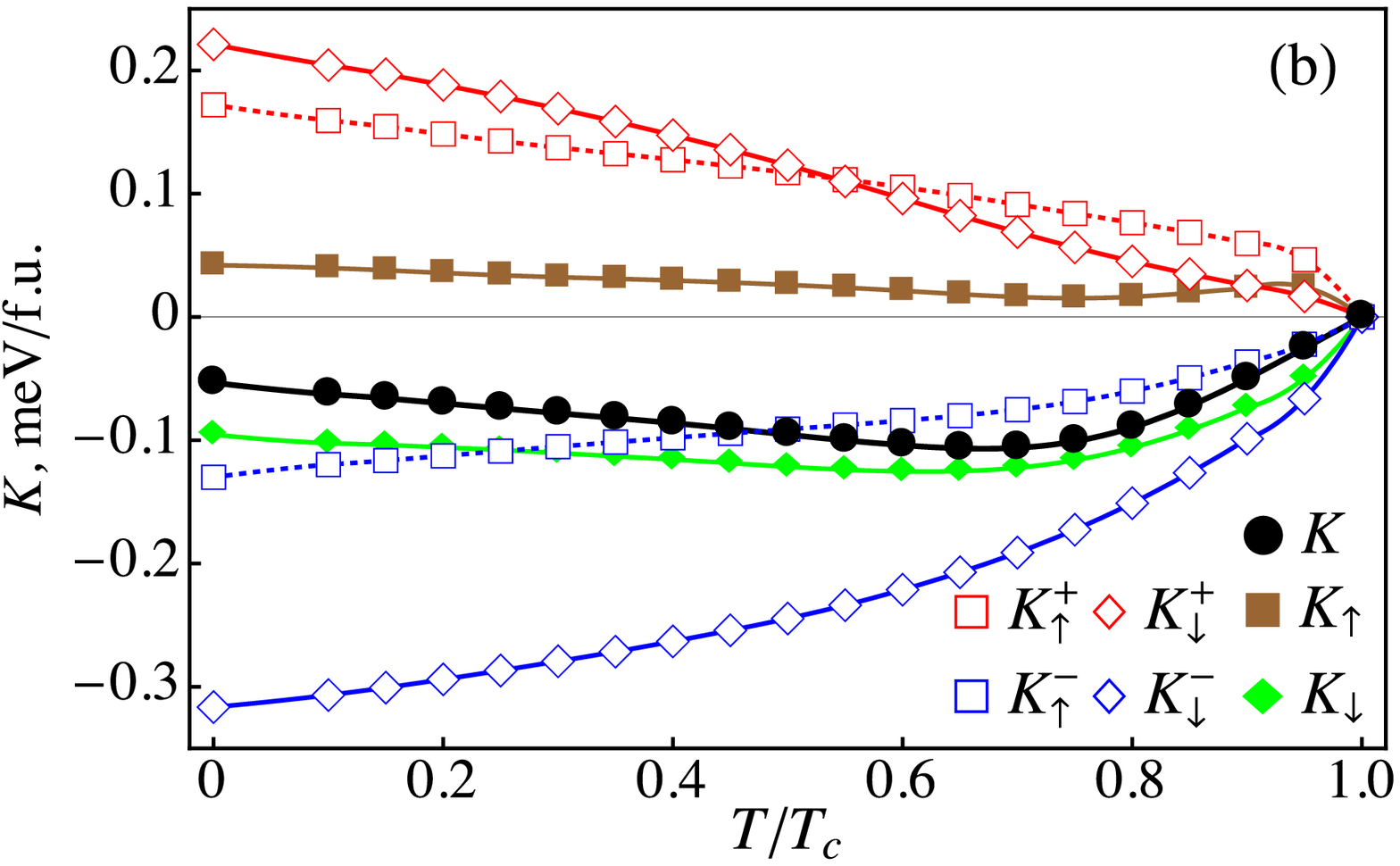}
\caption{Contributions to $K$ in (a) Fe$_2$B and (b) (Fe$_{0.05}$Co$_{0.95}$)$_2$B from different spins ($K_\uparrow$ and $K_\downarrow$). $K^+_{\sigma}$ and $K^-_{\sigma}$ in panel (b): total positive and negative contributions to $K_\sigma$ coming from different $\mathbf{k}$ points. (Dotted lines show $K^+_{\uparrow}$, $K^-_{\uparrow}$.)}
\label{fig:terms}
\end{figure}

Fig.\ \ref{fig:terms}a shows the temperature dependence of $K_\sigma$ in Fe$_2$B. Since the spin-mixing contribution $K_{\uparrow\downarrow}$ here is small (Fig.\ \ref{MAEvca}), $K_\uparrow$ and $K_\downarrow$ provide information similar to $K_{\uparrow\uparrow}$ and $K_{\downarrow\downarrow}$ at $T=0$ while retaining clear meaning at finite temperature \cite{SM}. We see that $K_\downarrow$ decreases quickly with increasing $T$. This happens because the downward shift and broadening of the minority-spin bands strongly suppress the negative minority-spin contribution to $K$. In contrast, the initial increase in $K_\uparrow$ mirrors the upward slope of $K_{\uparrow\uparrow}(x,0)$ as a function of $x$ \cite{APL}, which occurs as the majority-spin bands shift upward relative to $E_F$ with decreasing $\Delta$.
At elevated temperatures the majority-spin contribution becomes dominant, and $K$ undergoes an anomalous sign change, i.e., a spin-reorientation transition.

\begin{figure}[htb]
\includegraphics[width=0.9\columnwidth]{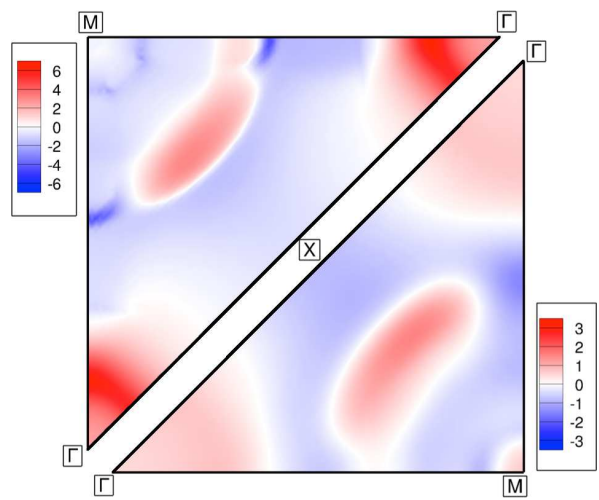}
\caption{Wave vector-resolved $K_\downarrow$ (units of meV$a_0^3$, where $a_0$ is the Bohr radius) on the $\Gamma$MX plane in (Fe$_{0.05}$Co$_{0.95}$)$_2$B alloy at $T=0$ (upper left) and $T/T_C=0.7$ (lower right).}
\label{fig:bz}
\end{figure}

At the Co-rich end the situation is complicated by the presence of large contributions of opposite sign that come from the minority-spin states in different regions of the Brillouin zone \cite{APL}. Near the $\Gamma$ point there is a large positive contribution from the degenerate bands that are mixed by $L_z$. There is also a large negative contribution from the mixing of minority-spin bands of opposite parity with respect to $\sigma_z$ reflection, which is distributed over the whole Brillouin zone. To help resolve these contributions, Fig.\ \ref{fig:terms}b for (Fe$_{0.05}$Co$_{0.95}$)$_2$B shows, in addition to $K_\sigma$, the total positive ($K^+_\sigma$) and negative ($K^-_\sigma$) contributions to $K_\sigma$, which were sorted by wave vector. Fig.\ \ref{fig:bz} displays $\mathbf{k}$-resolved $K_\downarrow$ on the $\Gamma MX$ plane at $T=0$ and $T/T_C=0.7$. The bright red ring around the $\Gamma$ point in Fig.\ \ref{fig:bz} is the hot spot coming from the two nearly-denegerate bands that are split by SOC (see Fig.\ \ref{fig:sf}a and \ref{splitting}).

As seen in Fig.\ \ref{fig:bz}, thermal spin disorder strongly suppresses the hot spot observed at $T=0$: it is strongly washed out at $T/T_C=0.7$, while the contributions from other regions decline almost homogeneously. This effect is similar to that of chemical disorder (Fig.\ \ref{splitting}). As a result, $K^+_\downarrow$ declines faster compared to other contributions shown in Fig.\ \ref{fig:terms}b, and the negative value of $K$ grows anomalously with $T$.

Interestingly, while in VCA the maximum in $K(x,0)$ with respect to band filling occurs near $x=0.95$ (Fig.\ \ref{MAEvca}), in CPA there is a cusped maximum \emph{exactly} in Co$_2$B. The latter is due to the fact that the bands are broadened by disorder with any admixture, reducing the positive contribution from the hot spots. This dominant effect of disorder explains why, as noted above, the anomalous $K(T)$ dependence at $x\geq0.9$ is opposite to the trend expected from increasing $x$, which holds at other concentrations. In Co$_2$B, where the positive contribution is at its maximum, both band broadening and decreasing $\Delta$ contribute to the anomalous decrease in $K(T)$, as the nearly degenerate bands broaden and sink below $E_F$.

In conclusion, we found that the anomalous temperature dependence of MCA in (Fe$_{1-x}$Co$_x$)$_2$B alloys is due to the changes in the electronic structure induced by spin fluctuations. This unconventional mechanism can be harnessed in applications where temperature-independent or increasing MCA is required.

\begin{acknowledgments}

The work at UNL was supported by the National Science Foundation through Grant No.\ DMR-1308751 and performed utilizing the Holland Computing Center of the University of Nebraska. Work at Ames Lab was supported in part by the Critical Materials Institute, an Energy Innovation Hub funded by the US DOE and by the Office of Basic Energy Science, Division of Materials Science and Engineering. Ames Laboratory is operated for the US DOE by Iowa State University under Contract No.\ DE-AC02-07CH11358.

\end{acknowledgments}

\balancecolsandclearpage

\setcounter{figure}{0}
\makeatletter
\renewcommand{\thefigure}{S\@arabic\c@figure}
\renewcommand{\bibnumfmt}[1]{[S#1]}
\renewcommand{\citenumfont}[1]{S#1}
\makeatother

\onecolumngrid
\begin{center}
\textbf{SUPPLEMENTAL MATERIAL}
\end{center}
\twocolumngrid

\section{Description of spin disorder}

\begin{figure}[htb]
\includegraphics[width=0.9\columnwidth]{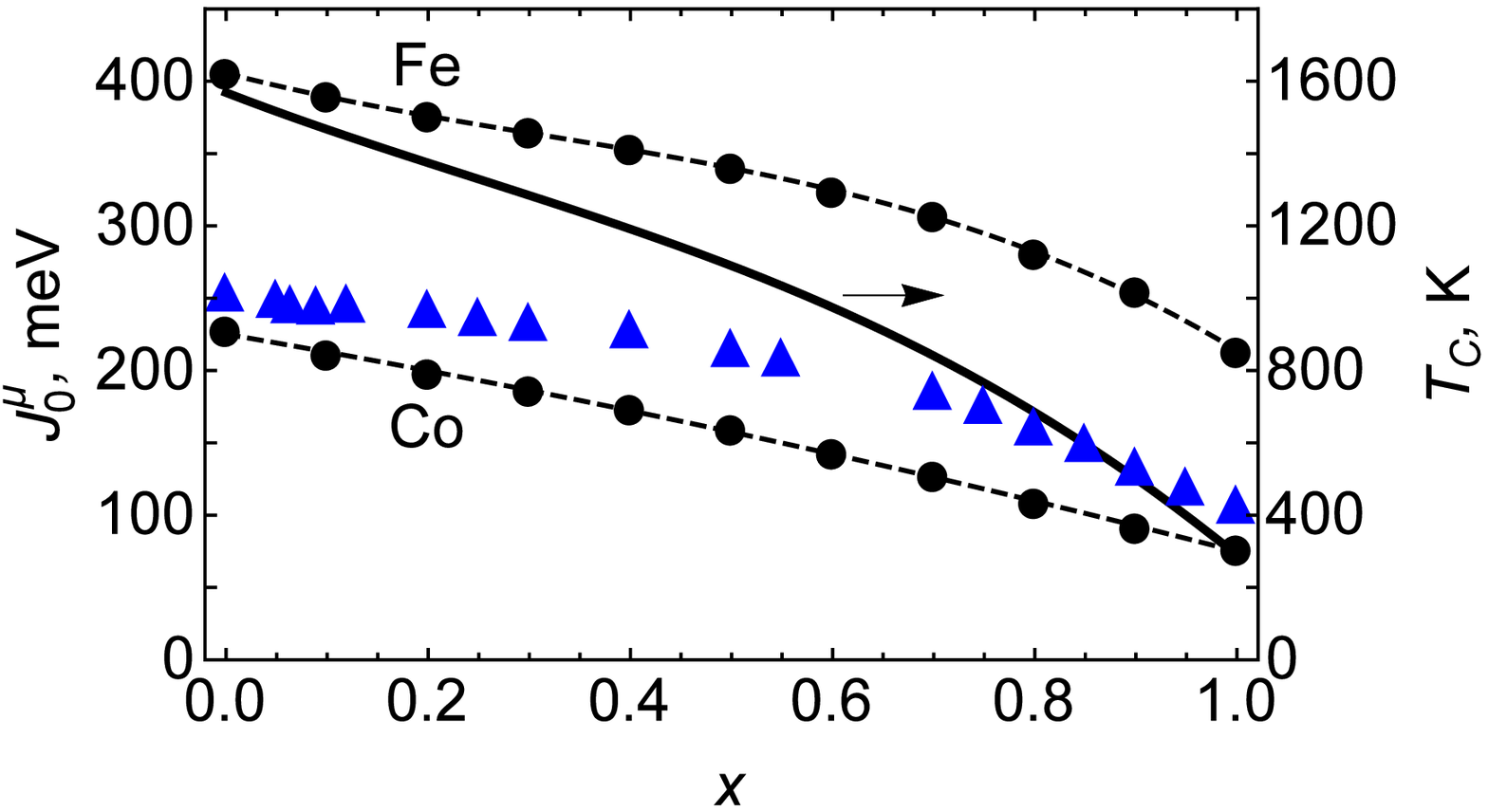}
\caption{Effective exchange parameters $J_{0}^{\mu}$ for Fe and Co (dashed lines) and calculated Curie temperature $T_C$ (solid line) as a function of concentration. Blue triangles: experimental data for $T_C$ \cite{S-Buschow}.}
\label{j0x}
\end{figure}

All calculations were performed using our implementation \cite{S-FeMnPt} of the disordered local moment (DLM) model for partially ordered magnetic states.
Thermal spin disorder was introduced as follows. Fig.\ \ref{j0x} shows the effective exchange parameters $J_{0}^{\mu}=\partial^2E/\partial\theta_\mu^2$ for both components ($\mu=\mathrm{Fe}$, Co) calculated at all concentrations using the linear response technique \cite{S-Liechtenstein}. These parameters describe the exchange interaction of an atom of a given type with the rest of the crystal. Within the mean-field approximation (MFA), these data predict the Curie temperatures $T_C$ of pure Fe$_2$B and Co$_2$B to be 1570 K and 290 K, which can be compared with experimental values of 1013 K and 429 K, respectively \cite{S-Buschow}.

For Fe$_2$B the $T_C$ is overestimated by about 35\%, which can largely be attributed to the neglect of short-range order in MFA. Indeed, linear response calculations show that the spin of a Fe atom is strongly ferromagnetically coupled to seven neighbors, five of which are within the layer. Although the couplings to more distant atoms are not negligible, they alternate in sign and contribute little to the effective exchange parameter. Thus, MFA can be expected to overestimate $T_C$ by about as much as it does (about 23\% \cite{S-Landau}) for the bcc lattice with nearest-neighbor exchange (coordination number 8). In contrast, for Co$_2$B the $T_C$ is underestimated by about a factor 1.5. This situation appears to be similar to the well-known case of fcc Ni, where $T_C$ calculated from $J_0$ is underestimated by nearly a factor of 2 \cite{S-Ni}. The similarity is due to the fact that both Ni and Co$_2$B are strongly itinerant ferromagnets with relatively small local moments. In such systems the effects of quantum spin fluctuations become significant, which for Co$_2$B is also reflected in the large overestimation of the magnetization in density-functional calculations. In addition, the long-wave approximation inherent in the linear-response calculation of $J_0$ becomes unreliable \cite{S-Antropov}.

For our present problem, it is important to capture the gradual increase of the spin disorder with increasing temperature leading to decreased exchange splitting and band broadening. It is likely that these effects are not sensitive to the moderate uncertainties in the calculation of $J_{0}^{\mu}$. The latter only affect the relative degree of disordering for Fe and Co, while the most interesting anomalies in $K(x,T)$ occur close to pure Fe$_2$B and Co$_2$B. Therefore, we adopt the following scheme based on the values of $J_{0}^{\mu}$ calculated above. The MFA equations for a two-component alloy described within the Heisenberg model contain four component-resolved parameters $J_{\mu\nu}$ defined as the exchange interaction of the spin of atom type $\mu$ with atoms of type $\nu$ everywhere else in the crystal. Introducing the pair exchange parameters $J_{\mu\nu}^{ij}$, we have $J_{\mu\nu}=x_{\nu}\sum_j J_{\mu\nu}^{ij}=x_{\nu}\tilde J_{\mu\nu}$ and $J_{0}^{\mu}=\sum_{\nu}J_{\mu\nu}$, where $x_{\nu}$ is the concentration of the component $\nu$. In our case this translates to $J_{0}^\mathrm{F}=(1-x)\tilde J_\mathrm{FF}+x\tilde J_\mathrm{FC}$ and $J_{0}^\mathrm{C}=(1-x)\tilde J_\mathrm{FC}+x\tilde J_\mathrm{CC}$, where we took into account that $\tilde J_\mathrm{FC}=\tilde J_\mathrm{CF}$ (F stands for Fe and C for Co).
We further fit the concentration dependence of $\tilde J_\mathrm{FC}$ and $\tilde J_\mathrm{CC}$ to a linear and $\tilde J_\mathrm{FF}$ to a quadratic polynomial in $x$ so as to best approximate the concentration dependence of $J_{0}^\mathrm{F}$ and $J_{0}^\mathrm{C}$ in Fig.\ \ref{j0x}. This fitting gives $\tilde J_\mathrm{FF}=406+220x^2$, $\tilde J_\mathrm{FC}=225-11x$, and $\tilde J_\mathrm{CC}=116-41x$ in meV units. (The linear term in $\tilde J_\mathrm{FF}$ is negligibly small.)

The coefficients $\alpha_\mu$ in the distribution functions for spin orientations are determined from the solution of the MFA matrix equation $T\bar{\alpha}=\tilde J \bar{m}$, where $\bar\alpha$ is a column vector with two elements $\alpha_\mu$, $\tilde J$ is a $2\times2$ matrix with elements $\tilde J_\mathrm{\mu\nu}$, and $\bar m$ is a column vector (of reduced component magnetizations) with elements $x_\mu L(\alpha_\mu)$, $L(\alpha)$ being the Langevin function. The Curie temperature $T_C$ corresponds to the vanishing of both $\alpha_\mu$. The MCA energy $K$ is calculated as a function of $x$ and $T/T_C$. To facilitate the comparison with experimental data, the $K(x,T)$ curves in Fig.\ 1 of the main text are plotted using the experimental values of $T_C$ for each concentration \cite{S-TAK}.

In the (Co$_{0.9}$Ni$_{0.1}$)$_{2}$B system the spin moments of Ni atoms are quite small, and in principle they should not be treated within the DLM model. However, such treatment does not introduce significant errors, because disorder of these small spin moments has a negligible effect on the electronic structure. Therefore, in this system we treated both Co and Ni within the DLM model with identical values of $\alpha$ for Co and Ni. In this way the system automatically tends to a paramagnetic state at $\alpha\to0$, and the self-consistent calculation of the local moment of Ni at each temperature is avoided.

\section{Spin-orbit coupling and magnetocrystalline anisotropy}

The spin-orbit coupling (SOC) is included as a perturbation of the potential parameters in the Green's function-based tight-binding linear muffin-tin orbital (LMTO) method \cite{S-APL,S-Kudr}, and the resulting band problem is solved exactly. This is analogous to the so-called ``pseudoperturbative'' treatment of SOC in conventional band-structure methods, in which the eigenvalues of the Hamiltonian perturbed by SOC are found exactly \cite{S-Andersen,S-Koelling,S-Solovyev}. The MCA energy is calculated as the difference in the single-particle energy for two directions of the magnetization, $\mathbf{M}\parallel x$ and $\mathbf{M}\parallel z$, calculated with the same potential parameters and distribution functions for spin orientations.

The coherent-potential equations involve an integration over the orientations of the spin local moment on each magnetic atom treated within the DLM model. For collinear magnetic orderings in the non-relativistic case the axial spin symmetry with respect to the direction of the magnetic order parameter is retained, and the integration over the azimuthal angle can be handled analytically \cite{S-FeMnPt}. SOC breaks this symmetry, and a full integration over the sphere needs to be taken. Here we used an 88-point quadrature for this integration, which was found to provide well-converged results. Thus, formally we apply the coherent potential approximation to an 176-component alloy (88 orientations of the local moment for Fe and Co). Each such component is described by LMTO potential parameter matrices which are first calculated in the reference frame of the local moment and then rotated to the prescribed direction by a rotation operator generated by the total angular momentum operator $\hat J$ \cite{S-FeMnPt}.

In the analysis of the underlying mechanisms of MCA, we employ two approximate decompositions of $K$. The first one utilizes the calculated anisotropy of the SOC energy:
\begin{equation}
K_{SO}=-\frac1{2\pi}\Delta \sum\limits_{\sigma\sigma'}\mathop{\mathrm{Im}}\int\limits^{E_F} V_{\sigma\sigma'}G_{\sigma'\sigma}dE \label{kso}
\end{equation}
Here and in the following we use the notation $\Delta A$ to denote the difference in the values of a quantity $A$ for magnetization oriented along $x$ and along $z$.
$V$ is the perturbing SOC operator, and $G$ the Green's function calculated with SOC. The quantity $K_{SO}$ is naturally separated
in four spin contributions $K_{\sigma\sigma'}$ using the identity $2\langle\mathbf{SL}\rangle=\langle L_{z'}\rangle_{\uparrow\uparrow}-\langle L_{z'}\rangle_{\downarrow\downarrow}+\langle L_+\rangle_{\downarrow\uparrow}+\langle L_-\rangle_{\uparrow\downarrow}$ \cite{S-APL}. The analysis of $K_{SO}$ is useful, because it approximates $K$ well unless second-order perturbation theory is strongly violated \cite{S-AKA}; in (Fe$_{1-x}$Co$_x$)$_2$B the concentration dependence of $K_{SO}$ agrees very well with that of $K$. This decomposition, however, loses its utility at finite temperatures.

The second decomposition is defined as follows. First, we can write the single-particle energy, calculated with SOC included, for magnetization direction $\mathbf{n}$ as
\begin{equation}
E^{\mathbf{n}}_{sp} = \int\limits^{E^{\mathbf{n}}_F}E N_\mathbf{n}(E)dE=E_F^0 Q_{val} + \int\limits^{E^{\mathbf{n}}_F}(E-E_F^0)N_\mathbf{n}(E)dE
\end{equation}
where $Q_{val}$ is the total valence charge, $N_\mathbf{n}(E)$ the density of states (DOS), and $E_F^0$ can be set to the value of the Fermi energy calculated without SOC or to $E_F^{\mathbf{n}}$ for some specific orientation of $\mathbf{n}$.
The replacement of $E^\mathbf{n}_F$ by $E_F^0$ in the upper limit of the last term introduces an error $\delta E_\mathbf{n}\sim N(E_F)(E^\mathbf{n}_F-E_F^0)^2$. We now define
\begin{equation}
K_\sigma=\int\limits^{E^0_F}(E-E^0_F)\Delta N_\sigma(E)dE ,\label{ksig}
\end{equation}
where $N_\sigma(E)=-\frac1\pi \mathop{\mathrm{Im}}\mathop{\mathrm{Tr}}G(E)$ is the DOS of spin $\sigma$. The sum $K_\uparrow+K_\downarrow$ differs from $K$ only in the term $\delta E_x-\delta E_z$, which we have verified to be negligible.

The definition (\ref{ksig}) is used explicitly in the calculations of $K_\sigma$. It is, however, useful to observe that in the perturbative regime there is a relation between $K_\sigma$ and $K_{\sigma\sigma'}$ at $T=0$. This can be seen by following the derivation in Ref.\ \onlinecite{S-APL} while sorting out the spin-dependent terms. The second-order correction to the density of states is
\begin{equation}
\delta N_\sigma(E)=-\frac 1\pi \mathop{\mathrm{Im}}\mathop{\mathrm{Tr}}G_{0\sigma} V G_0 VG_{0\sigma}
\end{equation}
where $V$ is the perturbing SOC operator and $G_0$ the spin-diagonal Green's function calculated without SOC.
Using the cyclic property of the trace and the relation $G^2_{0\sigma}\approx-\partial G_{0\sigma}/\partial E$, which in CPA is satisfied approximately as long as disorder is not too strong \cite{S-APL}, we obtain
\begin{equation}
K_\sigma= \sum\limits_{\sigma'}X_{\sigma\sigma'},\label{ks}
\end{equation}
where
\begin{equation}
X_{\sigma\sigma'}=\frac 1\pi \Delta\mathop{\mathrm{Im}}\mathop{\mathrm{Tr}}\int\limits^{E^0_F}(E-E^0_F)\frac{\partial G_{0\sigma}}{\partial E} V G_{0\sigma'} V dE.
\end{equation}
On the other hand, inserting the first-order correction for $G_{\sigma\sigma'}$ in (\ref{kso}), we have
\begin{align}
K_{\sigma\sigma'}\approx-\frac1{2\pi}\Delta \mathop{\mathrm{Im}}\int V_{\sigma\sigma'}G_{0\sigma'}V_{\sigma'\sigma}G_{0\sigma}dE. \label{Kss}
\end{align}
Note that, while it can be employed in a perturbative calculation of MCA energy \cite{S-Solovyev}, Eq.\ (\ref{Kss}) is not used explicitly here. Using integration by parts and ignoring the relatively small energy dependence of the SOC constants, it is now easy to show
\begin{equation}
K_{\sigma\sigma'}=\frac{X_{\sigma\sigma'}+X_{\sigma'\sigma}}{2} \label{kx}.
\end{equation}
Combining this with (\ref{ks}), we find, in the perturbative regime,
\begin{equation}
K_{\sigma}\approx K_{\sigma\sigma}+X_{\sigma\bar\sigma},\quad K_{\uparrow\downarrow}\approx\frac{X_{\uparrow\downarrow}+X_{\downarrow\uparrow}}{2},\label{krel}
\end{equation}
where $\bar\sigma\neq\sigma$. The relations (\ref{kx}) hold when the perturbative approximations are admissible, i.e., as long as the quasi-degenerate states near $E_F$ do not dominate in the MCA energy \cite{S-Solovyev,S-Kondorskii}.

If spin-off-diagonal band mixing, coming from the $L_\pm S_\mp$ terms in the SOC operator, can be neglected, then $K_\sigma\approx K_{\sigma\sigma}$ comes entirely from the mixing of states in spin channel $\sigma$ by the $L_zS_z$ operator, and $K_{\sigma\sigma}$ represents a perturbative approximation for $K_{\sigma}$. This correspondence holds at all concentrations where $K_{\uparrow\downarrow}$ is small, i.e., sufficiently far from $x=1$ \cite{S-APL}. However, even when $K_{\uparrow\downarrow}$ is appreciable, the reciprocal-space resolution of $K_{\sigma}$ can be used to identify the hot spots in reciprocal space and their origin. Indeed, mixing of nearly-degenerate bands of the same spin $\sigma$ near $E_F$ generates hot spots in $K_\sigma$, while strong off-diagonal band mixing should give rise to hot spots in both $K_\uparrow$ and $K_\downarrow$. The analysis of $K_\sigma$ naturally extends to finite temperatures.

\section{Comparison of $\bm{K(T)}$ with experiment}

\begin{figure}[hbt]
\includegraphics[width=0.9\columnwidth]{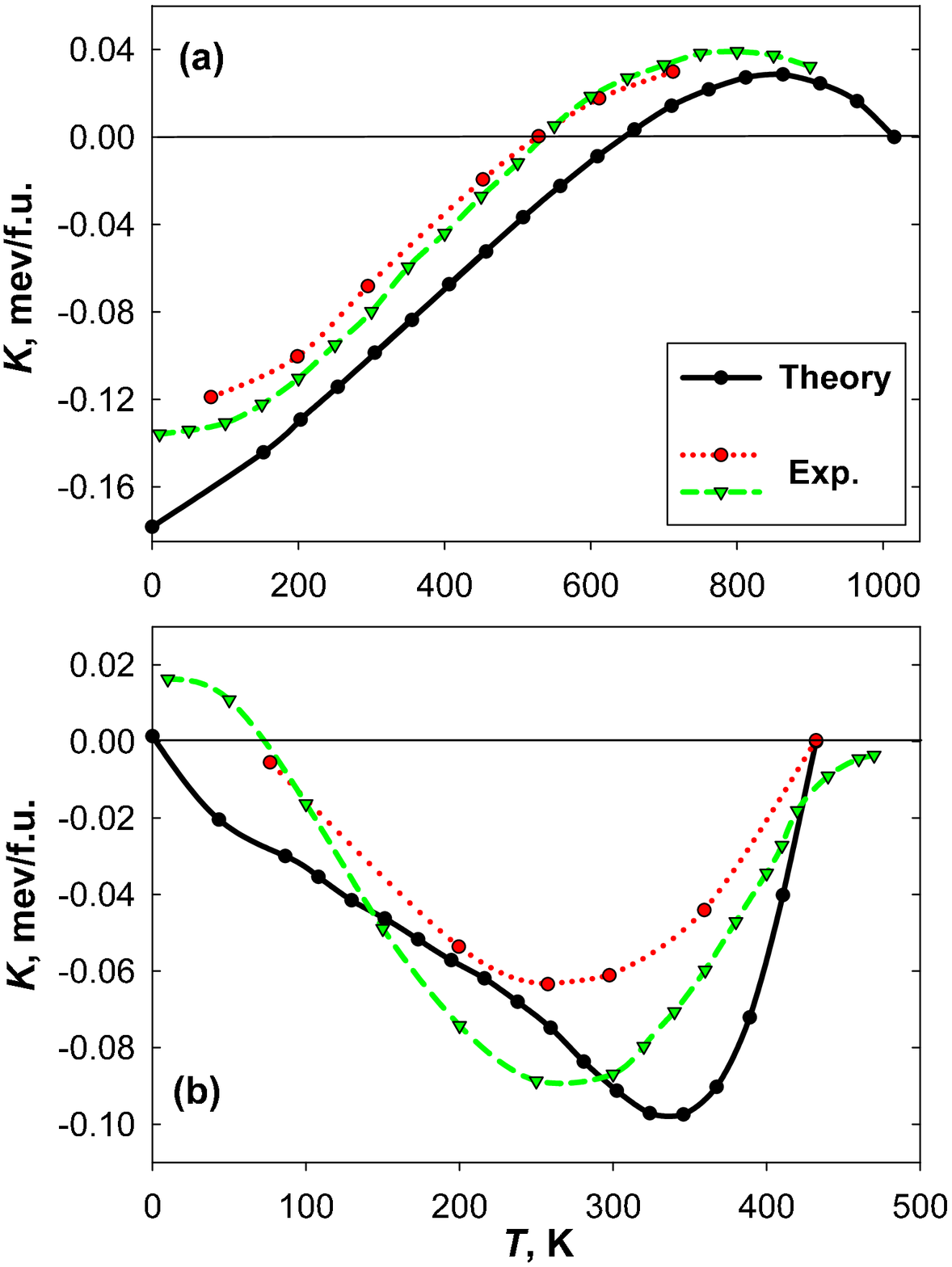}
\caption{Comparison of theoretical and experimental $K(T)$ curves for (a) Fe$_2$B, (b) Co$_2$B. Solid black lines: theory; red dotted lines: Ref.\ \onlinecite{S-Iga}; green dashed lines: Ref.\ \onlinecite{S-Edstrom}.}
\label{fig:ktexp}
\end{figure}

The calculated $K(T)$ curves for Fe$_2$B and Co$_2$B are compared with experimental data in Fig.\ \ref{fig:ktexp}. The experimental temperature dependence is well reproduced in the calculations. The differences between different experiments are discussed in Ref.\ \onlinecite{S-Edstrom}.

\section{Spectral functions for F\lowercase{e}$_2$B}

\begin{figure}[htb]
\vskip3ex
\includegraphics[width=0.9\columnwidth]{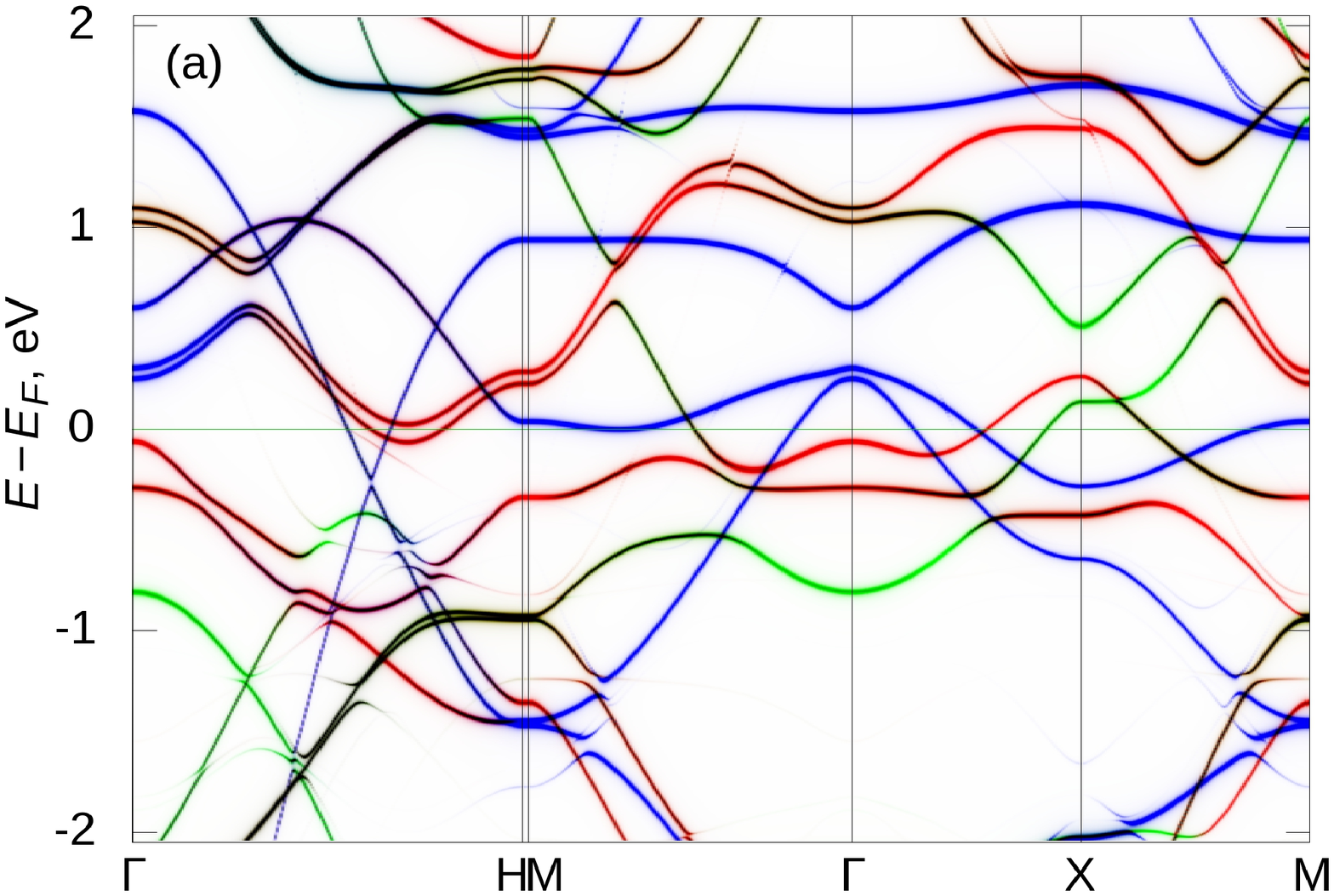}\vskip2ex
\includegraphics[width=0.9\columnwidth]{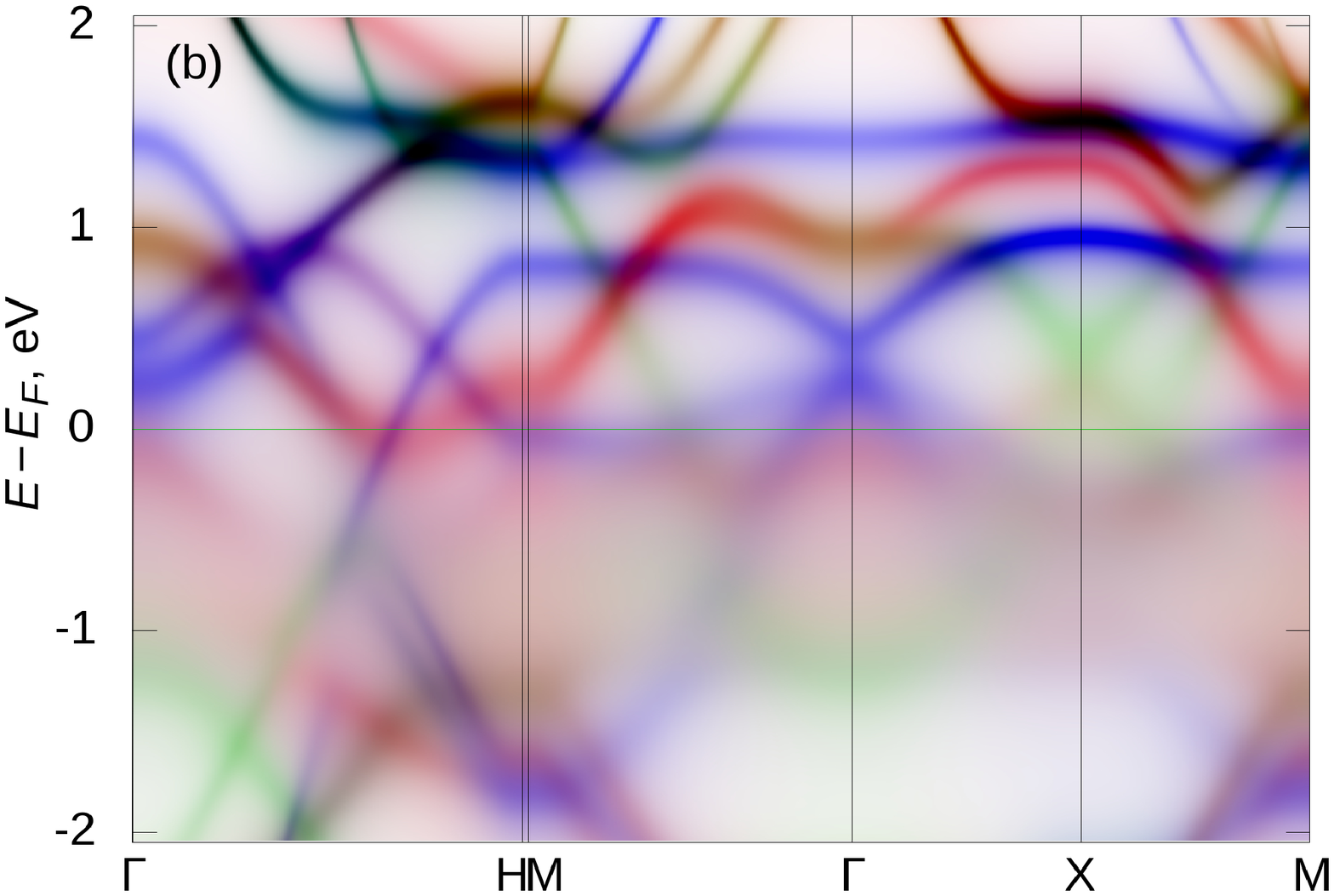}
\caption{Partial minority-spin spectral function for the transition-metal site in Fe$_2$B at (a) $T=0$, and (b) $T/T_C=0.7$. All details are as in Fig.\ 2 of the main text.}
\label{fig:sf2}
\end{figure}

Fig.\ \ref{fig:sf2} shows the Bloch spectral functions in Fe$_2$B at $T=0$ and $T/T_C=0.7$. These figures can be compared with Fig.\ 2 of the main text corresponding to the (Fe$_{0.05}$Co$_{0.95}$)$_2$B composition.
At the Co-rich end the dominant effect of spin disorder is the Stoner-like reduction of exchange splitting with only moderate band broadening. In contrast, at the Fe-rich end there is very strong band broadening, particularly in the 1 eV energy window below the Fermi level.

\section{Magnetocrystalline anisotropy in real space}

\begin{figure}[htb]
\includegraphics[width=0.9\columnwidth]{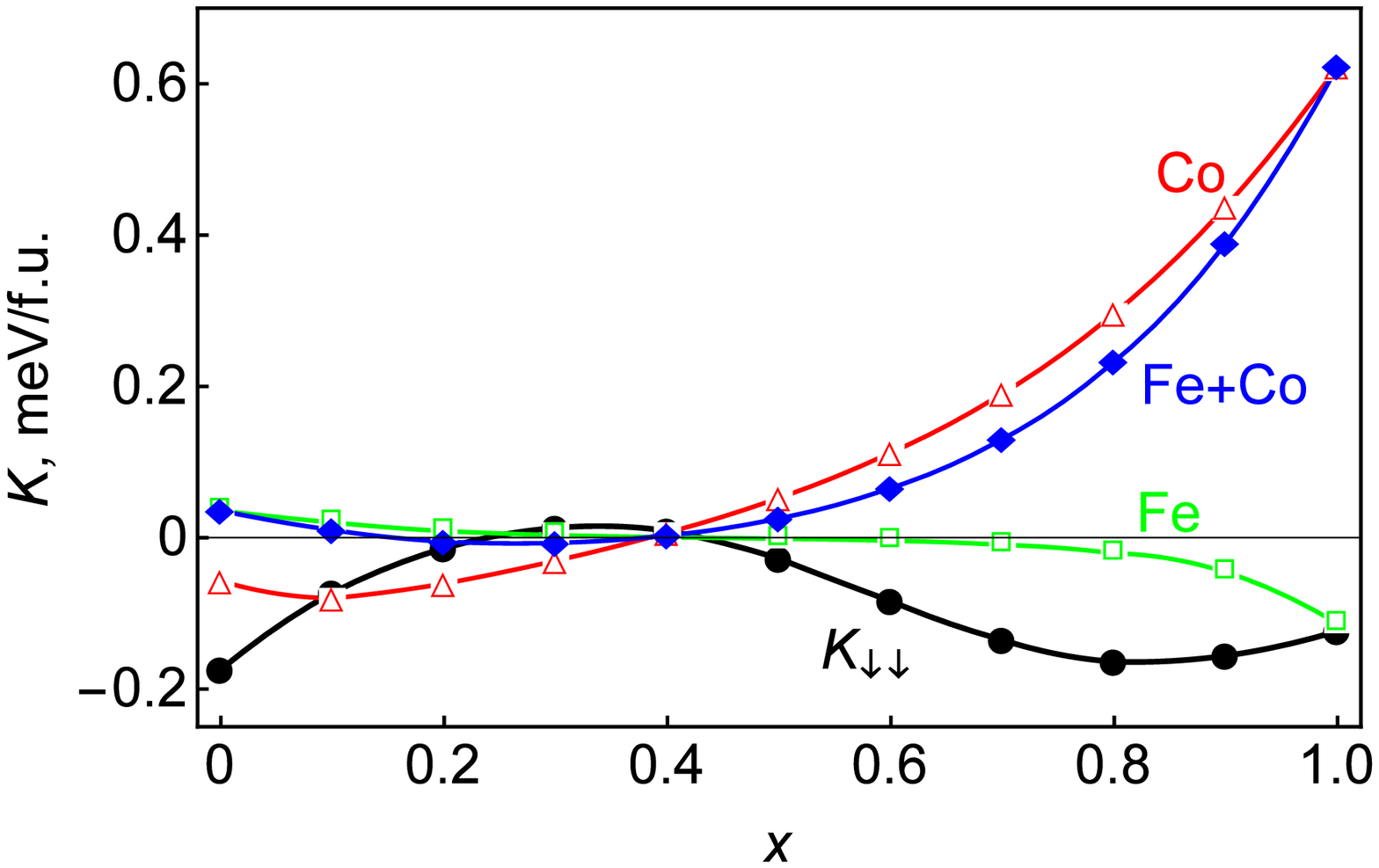}
\caption{Single-site terms in $K_{\downarrow\downarrow}$ (circles): two Fe sites (open squares), two Co sites (open triangles), and their concentration-weighted sum (filled diamonds).}
\label{ssite}
\end{figure}

The main text presents the analysis of MCA in reciprocal space. In principle, a real-space analysis could provide an alternative description. Fig.\ \ref{ssite} compares the $K_{\downarrow\downarrow}$ term with the estimated single-site contribution to it. The single-site terms for Fe (or Co) were computed by setting the SOC parameters to zero for all atoms except Fe (or Co) on one of the four transition-metal sites in the unit cell. If MCA were dominated by single-site terms, the concentration-weighted average of these terms would coincide with $K_{\downarrow\downarrow}$, but Fig.\ \ref{ssite} shows not even a correlation between them. Clearly, reciprocal space analysis is preferable to the real-space decomposition of MCA in this itinerant system.

\begin{thebibliography}{99}
\bibitem{Ni-note} The scaling of the exchange-correlation field in CPA is used only for Co. In VCA calculations the scaling factor is interpolated, similar to the nuclear charge.
\end{thebibliography}
\end{document}